\documentclass[twocolumn]{aastex63}
\received{\today}
\revised{\today}
\accepted{\today}
\submitjournal{ApJ}

\shorttitle{Tomography of the Jet in NGC\,2392}
\shortauthors{Guerrero et al.}
\graphicspath{{./}{figures/}}

\begin{document}

\title{Tomography of the unique on-going jet in the planetary nebula NGC\,2392}

\correspondingauthor{M.A.\,Guerrero}
\email{mar@iaa.es}

\author[0000-0002-7759-106X]{M.\,A.\,Guerrero}
\affil{Instituto de Astrof\'isica de Andaluc\'\i a, Glorieta de la Astronom\'\i a s/n, E-18008, Granada, Spain}

\author[0000-0002-7705-2525]{S.\,Cazzoli}
\affil{Instituto de Astrof\'isica de Andaluc\'\i a, Glorieta de la Astronom\'\i a s/n, E-18008, Granada, Spain}

\author[0000-0002-0121-2537]{J.\,S.\,Rechy-Garc\'\i a}
\affil{Instituto de Radioastronom\'ia y Astrof\'isica,  UNAM 
Campus Morelia, Apartado postal 3-72, 58090, Morelia, Michoacán, Mexico}

\author[0000-0003-2653-4417]{G.\ Ramos-Larios}
\affil{Instituto de Astronom\'\i a y Meteorolog\'\i a, CUCEI, 
Univ.\ de Guadalajara, 
Av.\ Vallarta 2602, Arcos Vallarta, 44130 Guadalajara, Mexico}

\author[0000-0001-9779-4895]{B.\,Montoro-Molina}
\affil{Instituto de Astrof\'isica de Andaluc\'\i a, Glorieta de la Astronom\'\i a s/n, E-18008, Granada, Spain}

\author[0000-0001-8252-6548]{V.M.A.\ G\'omez-Gonz\'alez}
\affil{Instituto de Radioastronom\'ia y Astrof\'isica,  UNAM 
Campus Morelia, Apartado postal 3-72, 58090, Morelia, Michoacán, Mexico}

\author[0000-0002-5406-0813]{J.\,A.\,Toal\'{a}}
\affil{Instituto de Radioastronom\'ia y Astrof\'isica,  UNAM 
Campus Morelia, Apartado postal 3-72, 58090, Morelia, Michoacán, Mexico}

\author[0000-0002-3981-7355]{X.\ Fang}
\affil{Key Laboratory of Optical Astronomy, National 
Astronomical Observatories, Chinese Academy of Sciences 
(NAOC), Beijing, China}
\affil{Department of Physics \& Laboratory for 
Space Research, Faculty of Science, 
University of Hong Kong, Hong Kong, China}




\begin{abstract}

Jets (fast collimated outflows) are claimed to be the main shaping agent of the most asymmetric planetary nebula (PNe) as they impinge on the circumstellar material at late stages of the asymptotic giant branch (AGB) phase. 
The first jet detected in a PN was that of NGC\,2392, yet there is no available image because its low surface brightness contrast with the bright nebular emission. 
Here we take advantage from the tomographic capabilities of GTC MEGARA high-dispersion integral field spectroscopic observations of the jet in NGC\,2392 to gain unprecedented details of its morphology and kinematics.  
The jet of NGC\,2392 is found to emanate from the central star, break through the walls of the inner shell of this iconic PN and extend outside the nebula's outermost regions with an S-shaped morphology suggestive of precession.  
At odds with the fossil jets found in mature PNe, the jet in NGC\,2392 is currently being collimated and launched. 
The high nebular excitation of NGC\,2392, which implies a He$^{++}$/He ionization fraction too high to be attributed to the known effective temperature of the star, has been proposed in the past to hint at the presence of a hot white dwarf companion.
In conjunction with the hard X-ray emission from the central star, the present-day jet collimation would support the presence of such a double-degenerate system where one component undergoes accretion from a remnant circumbinary disk of the  common envelope phase.

\end{abstract}



\section{Introduction}
\label{sec:intro}

Planetary nebulae (PNe) are shells of ionized gas surrounding the descendants of low- and intermediate-mass stars. 
The canonical Interacting Stellar Winds (ISW) model of PN formation \citep{Kwok1978,Balick1987} interpreted the round, elliptical and mildly bipolar morphology of PNe in terms of the interaction of two stellar winds: 
the current isotropic fast stellar wind and an asymptotic giant branch (AGB) dense slow wind with an equatorial density enhancement. 
The ISW model, however, fails for the most asymmetric PNe and those with multipolar morphology \citep{Balick2002}.  
Their formation rests upon a completely different paradigm based on the action of fast collimated outflows (i.e., jets) launched late in the AGB phase that impinge on the nebular envelope \citep{Sahai1998}.  
The collimation of these jets has been indeed confirmed to take place in the late AGB phase, when the heavy mass-loss of the central star feeds (or forms a circumbinary disk that feeds) an accretion disk around a main-sequence or sub-stellar companion \citep{Bollen2019}.

The origin of this paradigm shift dates back to the mid 80s, when it was discovered what was called {\it a high-velocity multi-knot bipolar mass flow} in NGC\,2392 \citep{Gieseking1985}. 
This first jet was followed by the detection of many symmetric pairs of knots or filamentary string of knots in PNe.  
As of today, almost 60 have been kinematically confirmed to be jets \citep[see][and references therein]{Guerrero2020}. 
Ironically, there is no direct image of the jet of NGC\,2392. 
Spatio-kinematic observations have confirmed its $\approx$ 200~km~s$^{-1}$ high velocity \citep{Reay1983,Balick1987,GarciaDiaz2012}, but the bright nebular emission of NGC\,2392 that overwhelms the emission of the jet, particularly at its bright inner shell, hinders the acquisition of a complete view of its morphology.  

The advent of high-dispersion integral-field spectroscopy (IFS) provides finally the means to resolve kinematically the emission of the jet of NGC\,2392 from that of the bright main nebula. 
High-dispersion IFS observations of NGC\,2392 have thus been acquired to provide a clean view of the extent, morphology and kinematics of its jet.  
Details of the instrument and data acquisition, reduction and analysis are described in \S2, the results are presented in \S3, and the implications are discussed in \S4.  
A short summary is given in \S5.

\section{OBSERVATIONS} 
\label{sec:observations}

\subsection{HST archive images}

\emph{Hubble Space Telescope} (\emph{HST}) Wide Field and Planetary Camera 2 (WFPC2) images of NGC\,2392 will be used through this work for different purposes.  
Images in the F502N, F656N, F658N, and F671N filters (program IDs 8499 and 8726) corresponding to the [O~{\sc iii}] $\lambda$5007 \AA, H$\alpha$ $\lambda$6563 \AA, [N~{\sc ii}] $\lambda$6584 \AA, and [S~{\sc ii}] $\lambda\lambda$6716,6731 \AA\ emission lines, respectively, were retrieved from the {\it Hubble} Legacy Archive\footnote{\url{https://hla.stsci.edu/}}.
The count rate of each image was divided by the total throughput efficiency of the filter at the line wavelength\footnote{
The information on the filter efficiency profile can be checked at \\
\url{https://www.stsci.edu/files/live/sites/www/files/home/hst/instrumentation/legacy/wfpc2/_documents/wfpc2_ihb.pdf}
to derive the surface brightness. 
The accuracy of these values is expected to be within $\approx$10\% as the filter profiles at the line wavelengths are notably flat \citep[e.g.,][]{BGR-L2021}.
}

\subsection{GTC MEGARA integral field spectroscopy}

IFS observations of NGC\,2392 were obtained on 2020 January 4 and 25 using the {\it Multi-Espectr\'ografo en GTC de Alta Resoluci\'on para Astronom\'\i a} \citep[MEGARA;][]{GildePaz2018} at the 10.4~m Gran Telescopio de Canarias (GTC). 
Both observing runs were performed under clear conditions and good seeing conditions with values in the range 1\farcs0-1\farcs2.
The high-resolution Volume-Phased Holographic (VPH) grism VPH665-HR was used, providing a spectral dispersion of 0.098~\AA~pix$^{-1}$ and a full-width at half-maximum (FWHM) spectral resolution $\approx$ 16~km~s$^{-1}$, that is, $R\approx18,700$.
The spectral range 6405.6--6797.1~\AA\ covered by the VPH665-HR grism includes the key emission lines of [N\,{\sc ii}]~$\lambda\lambda$6548,6584~\AA, H$\alpha$, and [S\,{\sc ii}]~$\lambda\lambda$6716,6731~\AA.

\begin{figure}
\begin{center}
\includegraphics[width=1.00\linewidth]{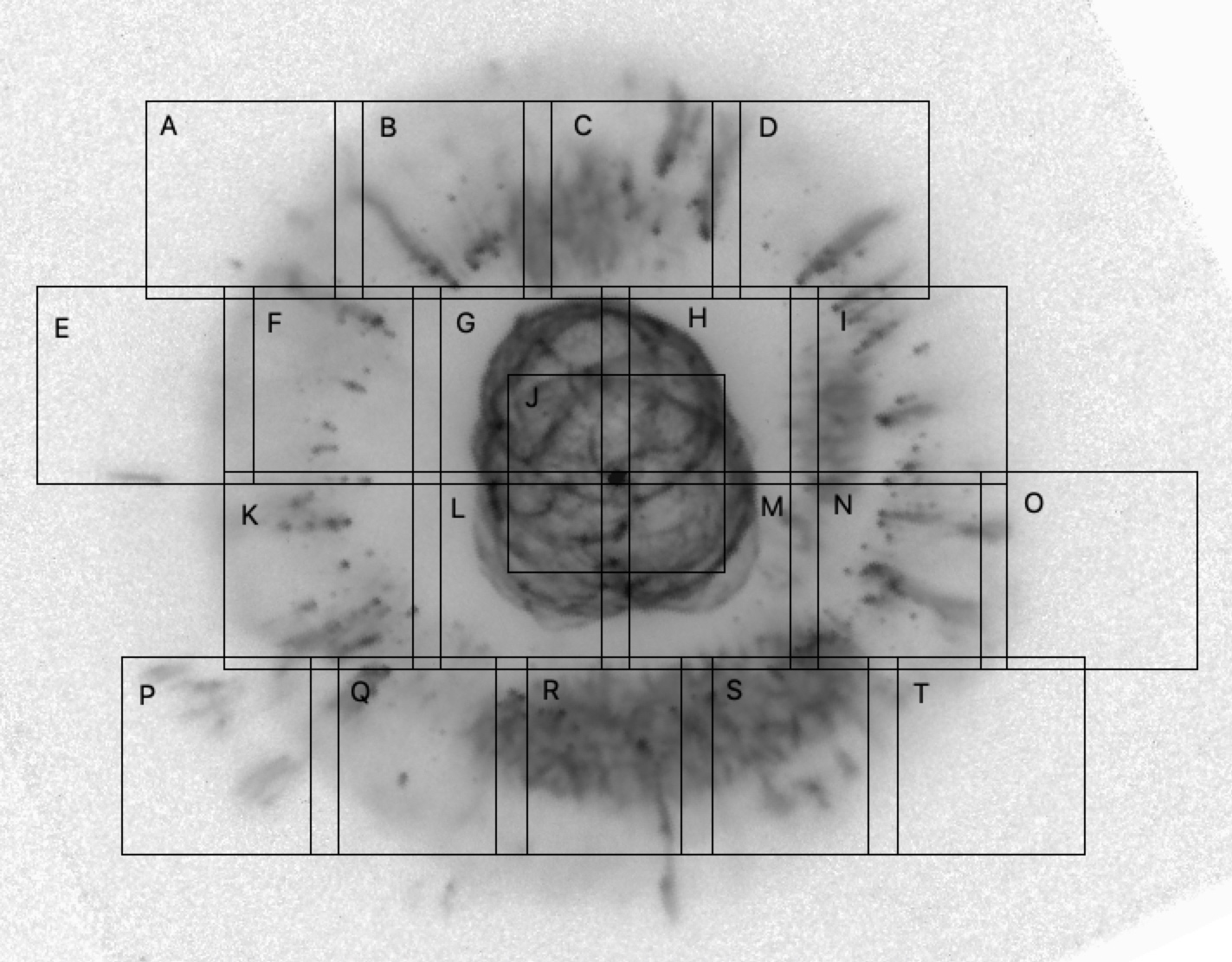} 
\caption{
\emph{HST} WFPC2 image of NGC\,2392 in the [N~{\sc ii}] $\lambda$6584 \AA\ emission line overlaid with the locations of the 20 MEGARA IFU fields.
The field of view of the MEGARA IFU is 12\farcs5$\times$11\farcs3.  
North is up, east to the left.  
}
\vspace*{-0.35cm}
\label{fig:ifu}
\end{center}
\end{figure}

The integral-field unit (IFU) mode was used. 
It has 567 hexagonal spaxels with a diameter of 0\farcs62 and a field of view (FoV) of 12\farcs5$\times$11\farcs3. 
Since the nebular size of NGC\,2392 ($\sim48''$) is larger than the instrument FoV, observations in twenty different pointings were obtained as shown in Figure~\ref{fig:ifu} to cover the inner shell and most of the outer nebular shell, with special coverage of the diameter along position angle (PA) 70$^\circ$ to map the outflow \citep[see][]{GarciaDiaz2012}, which is detected in pointings ``A'', ``B'', ``E'' to ``O'', ``S'' and ``T''. 
The FoV of nineteen out of the twenty pointings overlap $\sim2''$ along the East-West direction and 0\farcs5 along the North-South direction with adjacent fields.  
The twentieth pointing (``J'' in Fig.~\ref{fig:ifu}) was centered at the location of the central star. 
Three 5-minute exposures were obtained at each position to facilitate removal of cosmic rays.

\subsubsection{Data reduction}

The MEGARA raw data were reduced following the Data Reduction Cookbook using
the {\it megaradrp} v0.10.1 pipeline released on 2019 June 29 \citep{Pascual2019}. 
The data were sky and bias subtracted and flat-fielded. The sky subtraction used 56 ancillary fibers located $\approx2\farcm0$ from the IFU center. 
The individual spectra from each spaxel were then traced, extracted and wavelength calibrated to make a row-stacked spectrum (RSS). 
The RSS was then converted into $52\times58\times4300$ data cubes with 4300 elements in the spectral direction and $52\times58$ arrays of 0\farcs2 square spaxels in the spatial dimension using the pipeline regularization grid task {\it megararss2cube}. 
The data were subsequently flux calibrated using observations of spectro-photometric standard stars.

The data cubes were then ``sliced'' to produce maps in the H$\alpha$, [N~{\sc ii}] $\lambda$6584 \AA, and [S~{\sc ii}] $\lambda\lambda$6716,6731 \AA\ emission lines.  
The surface brightness of these maps were matched with those in the available \emph{HST} images in the corresponding F656N, F658N, and F671N filters applying second order flux-calibration corrections specific to each MEGARA IFU field.  
This procedure guarantees that the relative uncertainties in the flux-calibration among different fields are minimized. 
Furthermore, the comparison of the location of nebular features in the MEGARA maps and \emph{HST} images allowed us to derive precise spatial offsets between the MEGARA maps. 

\begin{figure}
\begin{center}
\includegraphics[bb=165 365 390 730,width=0.95\linewidth]{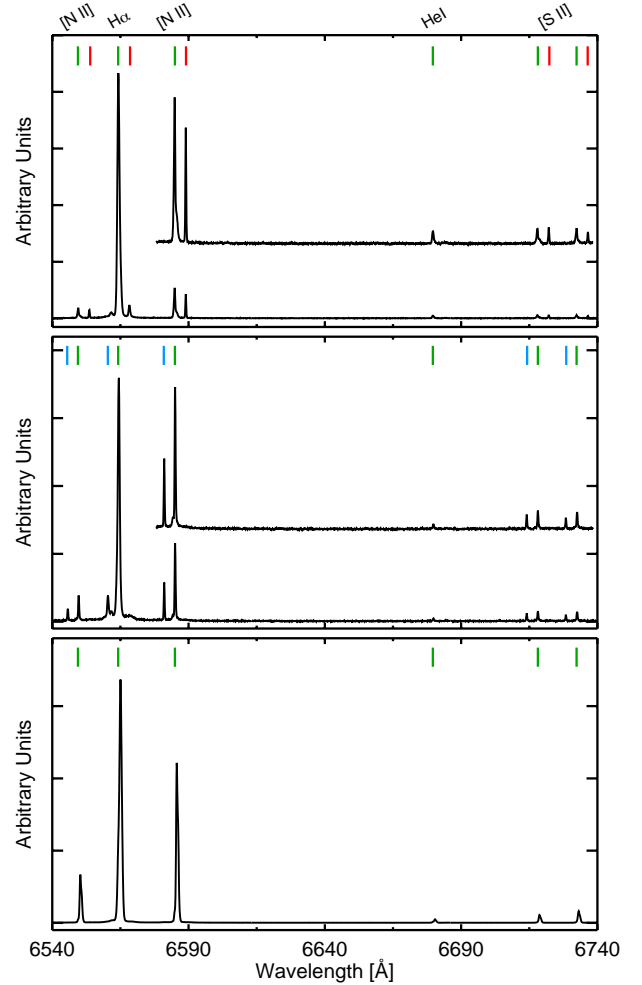} 
\caption{ 
Examples of spectra of individual spaxels showing the receding jet (top), approaching jet (middle), and a nebular region offset from the jet (bottom). 
The vertical green lines mark the systemic rest wavelength, whereas the vertical red and blue lines mark the receding and approaching components of the jet, respectively. 
In the top and middle panels, the spectral region covering the [N~{\sc ii}] $\lambda$6584 \AA\ and [S~{\sc ii}] emission lines of the spaxels registering jet emission is also displayed at a different intensity level to show more clearly the faintest spectral features. 
}
\vspace*{-0.35cm}
\label{fig:spec_all}
\end{center}
\end{figure}

\begin{figure*}
\begin{center}
\includegraphics[width=0.9\linewidth]{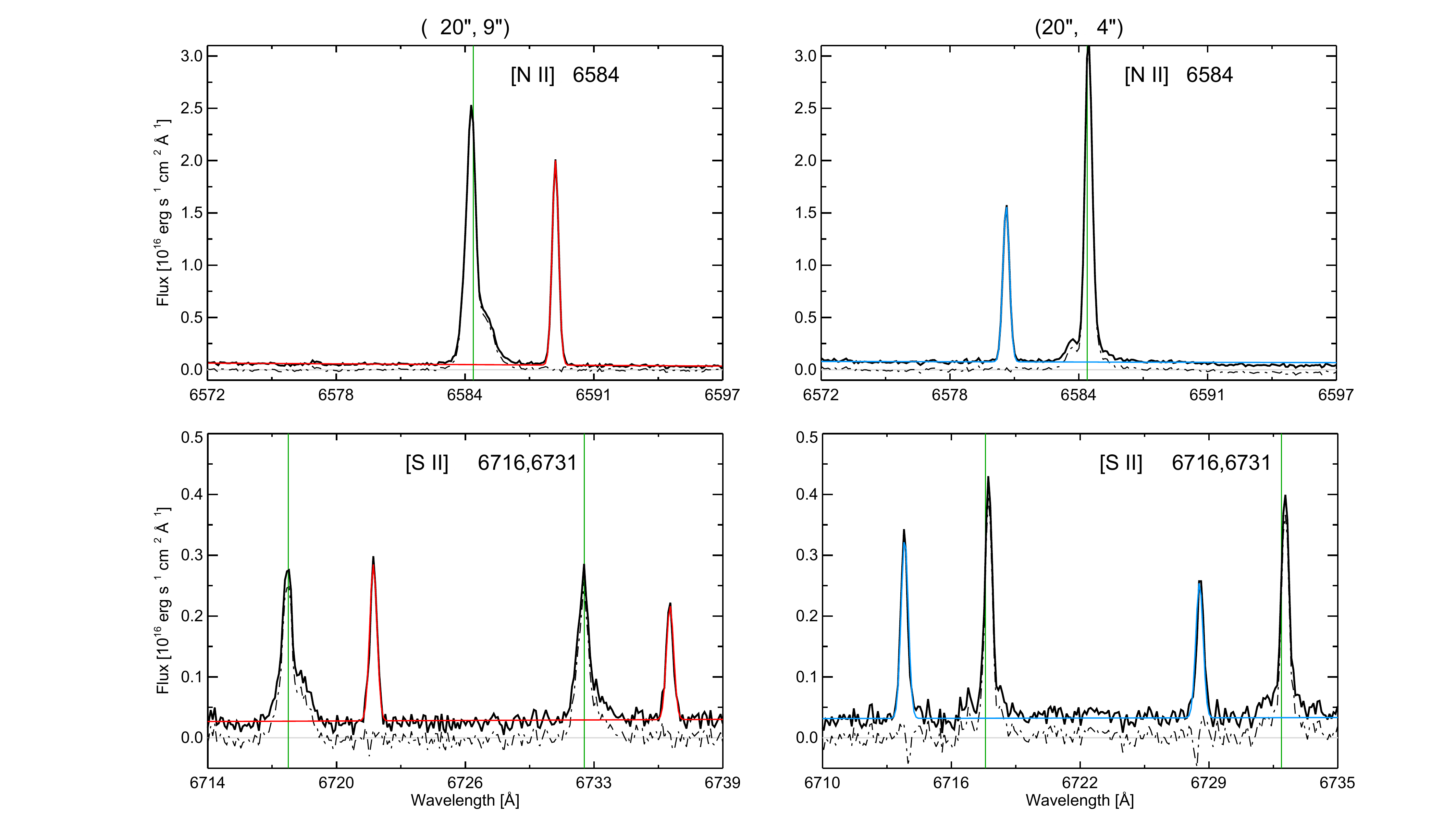} 
\caption{
Examples of fits to the spectra of individual spaxels using a single Gaussian function to the [N\,{\sc ii}] $\lambda$6584 \AA\ (top) and [S\,{\sc ii}] $\lambda\lambda$6716,6731 \AA\ (bottom) emission lines from the receding (left) and approaching (right) components of the jet of NGC\,2392. 
The fits are shown with red and blue lines for the receding and approaching components, respectively. 
The vertical green lines mark the systemic rest wavelength and are coincident with the emission from the nebula. 
The residuals of the fits are shown using dashed lines.
The spatial offsets of the spectra with respect to the central star ($\Delta\alpha,\Delta\delta$) are labeled at the top.
}
\vspace*{-0.35cm}
\label{fig:spec}
\end{center}
\end{figure*}

\subsubsection{Data analysis: mapping the jet}

As a first step in the data analysis, the MEGARA data cubes were ``sliced'' at a particular wavelength range to produce a pseudo-narrow-band image at a specific velocity interval of an emission line. 
This provides only crude information of the spatial distribution and kinematics (radial velocity and velocity width) of the gas in the jet of NGC\,2392.  
An inspection of the H$\alpha$, [N~{\sc ii}] and [S~{\sc ii}] emission line profiles at different locations reveals the presence of multiple kinematic components associated either with the nebula (bottom panel of Fig.~\ref{fig:spec_all}) or with both the nebula and the jet (middle and top panels of Fig.~\ref{fig:spec_all}). 
The emission of the high-velocity jet is generally well isolated from that of the nebular shells in the [N~{\sc ii}] and [S~{\sc ii}] emission lines. 
Thus the line intensities, centroids and widths at each spaxel can be fitted using a single Gaussian function.
On the other hand, the H$\alpha$ jet emission cannot be resolved properly from the H$\alpha$ nebular emission at all locations given its low surface brightness contrast with the H$\alpha$ nebular emission. 
These difficulties are further aggravated by the broader thermal width of this line.
Finally, we note that the [N~{\sc ii}] $\lambda$6548 \AA\ line can be deprecated in favor of the three-times-brighter [N~{\sc ii}] $\lambda$6584 \AA\ line.

Taking into account all comments above, we applied a procedure to obtain the best single-Gaussian fit of the [N~{\sc ii}] $\lambda$6584 \AA\ and [S~{\sc ii}] emission line profiles on a spaxel-by-spaxel basis (see Figure~\ref{fig:spec} for examples of these fits). 
The fitting was performed to all spaxels with sufficient signal-to-noise ratio using a the Levenberg-Marquardt least-squares fitting routine {\sc mpfitexpr} \citep{Markwardt2009} within the Interactive Data Language\footnote{http://www.harrisgeospatial.com/SoftwareTechnology/IDL. aspx} (IDL) environment for the [N~{\sc ii}] $\lambda$6584 \AA\ and [S~{\sc ii}] emission lines separately \citep[e.g.,][]{Cazzoli2020}. 
We found that constraints to the line widths of FWHM $<1$ \AA\ ($\approx$ 45 km~s$^{-1}$ at the rest-frame wavelengths of these lines) and absolute radial velocities with respect to the systemic velocity\footnote{
The systemic radial velocity, $v_{\rm sys}$, was found to be 70.5 km~s$^{-1}$, in agreement with the recent determination presented by \citet{GarciaDiaz2012}.
} 
($|v_r|$) between 130 and 250 km~s$^{-1}$ produced optimal results.  
At each spatial position, we additionally imposed that each [S~{\sc ii}] emission line shares the same kinematics (velocity and velocity dispersion).

\begin{figure}
\begin{center}
\includegraphics[width=1.02\linewidth]{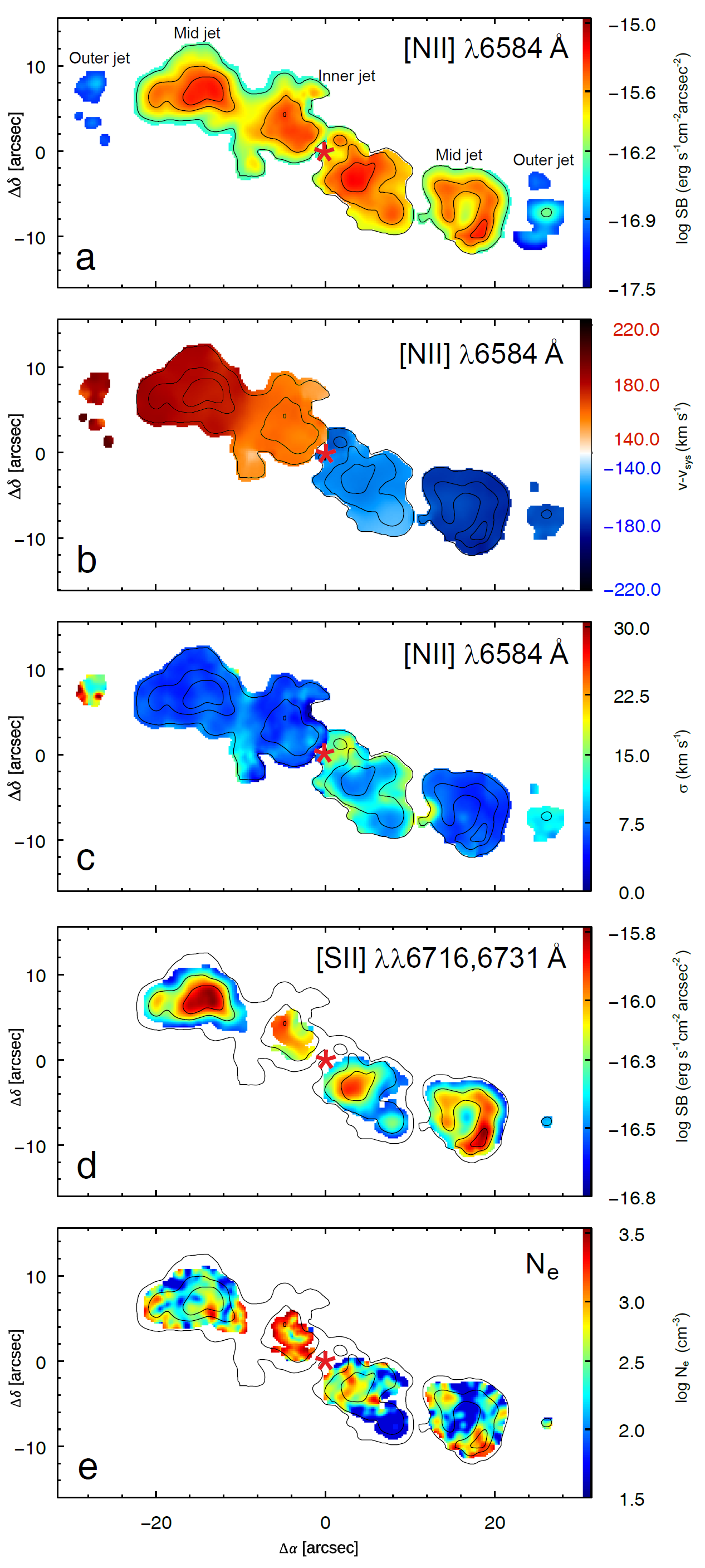} 
\caption{
GTC MEGARA maps of the jet of NGC\,2392. 
The different panels show the [N~{\sc ii}] $\lambda$6584 \AA\ 
(a) surface brightness (SB), 
(b) radial velocity with respect to the systemic velocity $v-v_{\rm sys}$ and 
(c) velocity width $\sigma$, 
(d) the [S~{\sc ii}] $\lambda\lambda$6716,6731 \AA\ surface brightness, and 
(e) the electronic density $N_e$. 
The contours of the [N~{\sc ii}] $\lambda$6584 \AA\ surface brightness map are overlaid over all maps and the location of the central star is marked by a red ``$\star$'' symbol. 
The location of the inner, mid and outer jets are marked on panel a.
}
\vspace*{-0.35cm}
\label{fig:megara}
\end{center}
\end{figure}

The output consists of fit parameters for each spectral feature including the central wavelength, width and flux of the line along with their uncertainties. 
A heliocentric velocity correction of $\sim$ 4 km~s$^{-1}$ was applied (the exact value depending on the observing time of each pointing). 
Finally, the observed line width was corrected for the effect of instrumental width ($\sigma_{\rm ins} \approx 6.8$ km~s$^{-1}$) by subtracting it quadratically from the observed line width.
The final maps of surface brightness, radial velocity with respect to the systemic velocity and instrumental-width-corrected velocity width of the jet are presented in Figure~\ref{fig:megara}.  
In these maps, regions for which the significance of the [N~{\sc ii}] $\lambda$6584 \AA\ and [S~{\sc ii}] emission lines is lower than 5$\sigma$ and 3$\sigma$, respectively, have been masked out. 
The kinematics in the [S~{\sc ii}] emission lines, which is basically consistent with that of the [N~{\sc ii}] emission line, is not presented.

\section{Results} 
\label{sec:results}

\subsection{Jet morphology, kinematics and density}

The jet in NGC\,2392 consists mostly of two blobs of similar angular extent and a few fainter outer knots (Fig.~\ref{fig:megara}-a). 
These three components will be referred to hereafter as the inner, mid and outer jets, respectively. 
Despite the morphological differences between the Western and Eastern components, the three components of the jet appear to twist point-symmetrically from the central star in a subtle S-shaped morphology. 
The total flux in the [N~{\sc ii}] $\lambda$6584 \AA\ emission line is 6.6$\times$10$^{-14}$ erg~cm$^{-2}$~s$^{-1}$, i.e., approximately 1000 times fainter than the nebula emission in this line. 
The average (peak) surface brightness of the jet in this line is 1.5$\times$10$^{-16}$ (4.5$\times$10$^{-16}$) erg~cm$^{-2}$~s$^{-1}$~arcsec$^{-2}$, with a jet-to-nebula surface brightness ratio peaking at $\approx$ 0.5 for the mid jet, but staying lower than 0.01 for the inner jet that is projected onto the bright inner shell.

\begin{figure*}
\begin{center}
\includegraphics[width=0.455\linewidth]{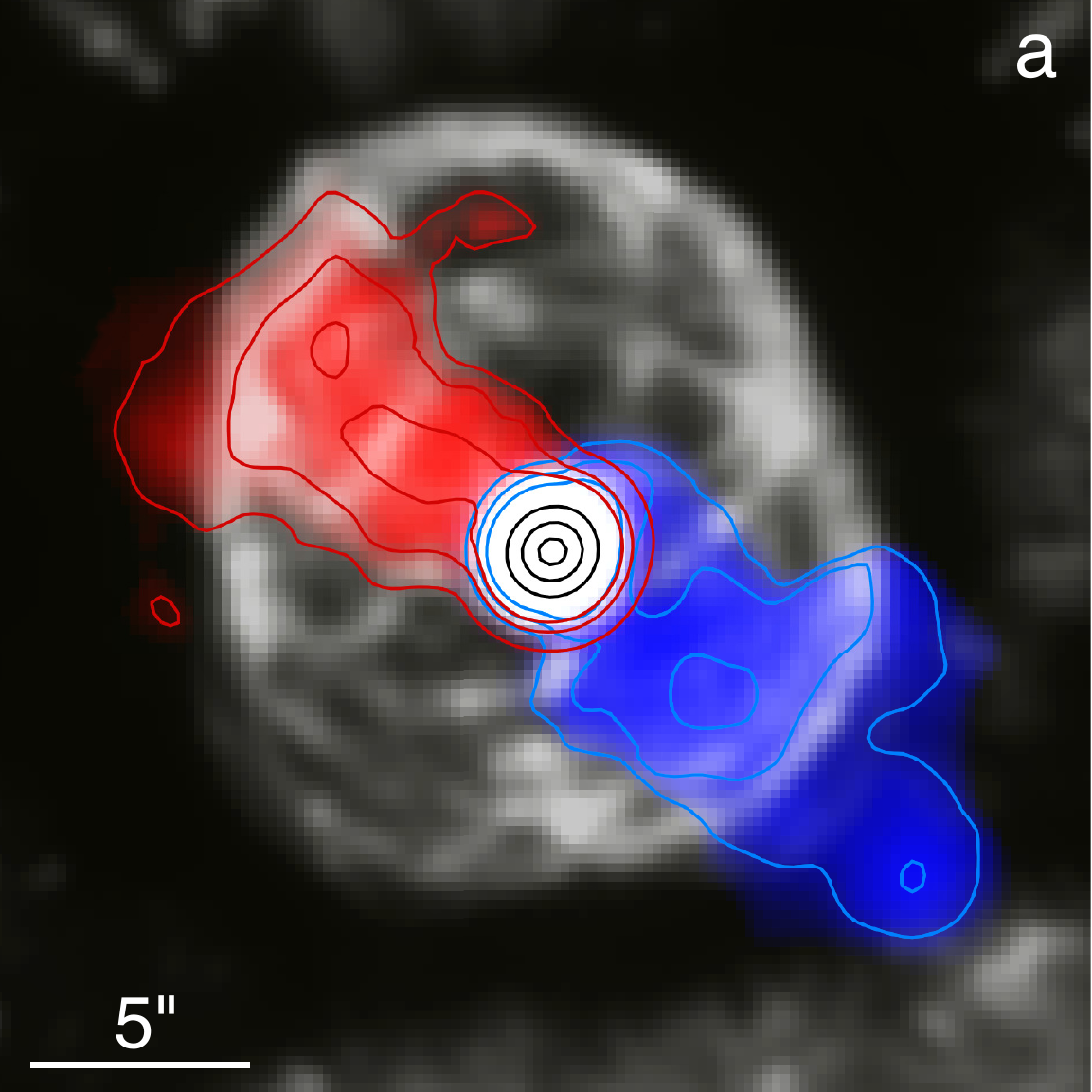} 
\hspace*{0.012cm}
\includegraphics[width=0.525\linewidth]{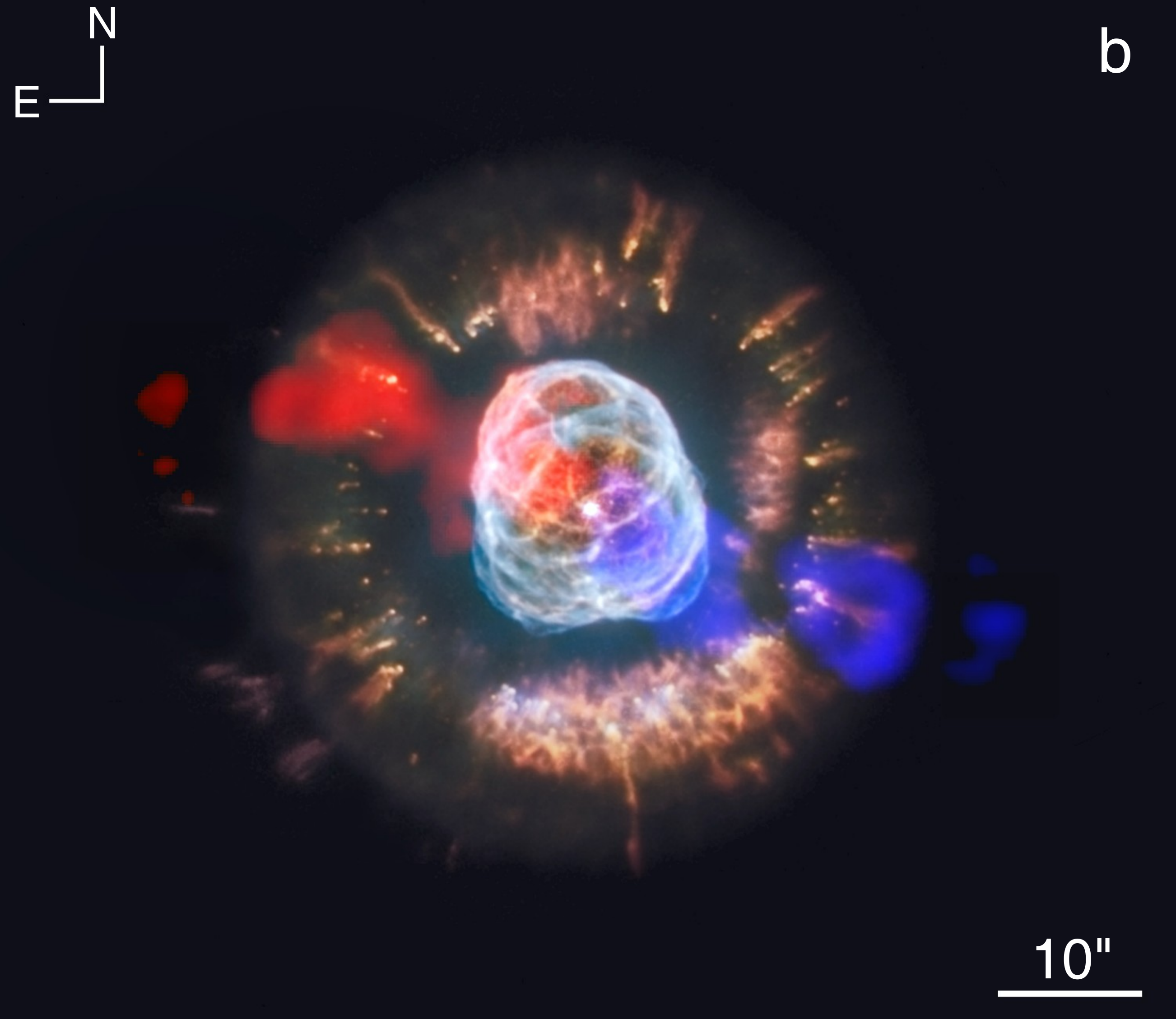} 
\caption{
Color-composite pictures of the nebular and jet emission of NGC\,2392. 
(left) GTC MEGARA images of the inner shell (white) and inner jet (red and blue) with pseudo-filters in the [N~{\sc ii}] $\lambda$6584 \AA\ emission line in the velocity range of the approaching (--190 to --150 km~s$^{-1}$, blue) and receding (+150 to +190 km~s$^{-1}$, red) components of the jet and nebula (--135 to +135 km~s$^{-1}$, white).
The blue and red contours emphasize the emission from the approaching and receding components of the jet, respectively. 
(right) \emph{HST} WFPC2 F502N, F656N and F658N color-composite picture of NGC\,2392 and GTC MEGARA [N~{\sc ii}] $\lambda$6584 \AA\ image of its jet (Fig.~\ref{fig:megara}-a). 
The approaching and receding components of the jet are depicted in blue and red, respectively.
}
\vspace*{-0.35cm}
\label{fig:img}
\end{center}
\end{figure*}

The comparison of the jet morphology and spatial extent with the nebular emission shown in Figure~\ref{fig:img} is clarifying. 
In spite of the much brighter nebular emission, the inner jet can be traced down to exactly to the location of the central star of NGC\,2392 (Fig.~\ref{fig:img}-left).
The inner jet brightens then at a location immediately interior to the inner shell rim and breaks through the walls of the inner shell reaching the nebular outer shell (Fig.~\ref{fig:img}-left).  
Meanwhile, the mid jet extends across the nebula outer shell, with its outermost tips being found outside the nebular edge (Fig.~\ref{fig:img}-right). 
The outer jet is definitely beyond the nebular outermost edge.

The velocity map in Figure~\ref{fig:megara}-b shows the Western component of the jet to approach and the Eastern component to recede from us. 
The jet velocity varies steadily with radius, with a notable $\approx$ 20 km~s$^{-1}$ increase from the inner to the mid jet and a small decrease between the mid and outer jets. 
The velocity breaks among the different components of the outflow are suggestive of episodic ejections, with small changes in the direction of the motion of the different components of the outflow.
On the other hand, the velocity dispersion in Figure~\ref{fig:megara}-c is notably small, close to the thermal width of the line at an electronic temperature ($T_e$) of 10,000 K. 
Thus, although the varying morphology of the approaching and receding components of the jet hints at a complex dynamics of the outflow in its interaction with the nebula, the coherent kinematics rather implies a steady motion of the gas in each component of the jet with little interaction with the bulk of the nebular material.

The GTC MEGARA high-dispersion IFS observations presented in panels a, b and c of Figure~\ref{fig:megara} provide the first clean and complete view of the morphology and kinematics of the jet of NGC\,2392 in the [N~{\sc ii}] $\lambda$6584 \AA\ emission line.  
The spatial extent of the inner jet could not be described in previous works based on Fabry-Perot \citep{Reay1983} and systematic mappings using multi-long-slit echelle spectroscopic \citep{Balick1987,GarciaDiaz2012} observations due to the extremely low emission contrast between the inner jet and the inner shell.  
That issue is further aggravated by the unexpectedly \citep{Jacob2013} fast expansion velocity of this shell \citep[$\approx$120 km~s$^{-1}$,][]{GarciaDiaz2012}, which makes difficult to deblend the emission of the jet from that of the inner shell. On the other hand, the outer jet could not be detected in previous observations due to its low surface brightness.

The morphology of the jet in the [S~{\sc ii}] $\lambda\lambda$6716,6731 \AA\ emission lines (Fig.~\ref{fig:megara}-d) is very similar to that in [N~{\sc ii}] although the total flux in the [S~{\sc ii}] lines is 1.3$\times$10$^{-14}$ erg~cm$^{-2}$~s$^{-1}$, i.e., about 5 times fainter. 
The average (peak) surface brightness of the [S~{\sc ii}] emission lines is 5.7$\times$10$^{-17}$ (1.4$\times$10$^{-16}$) erg~cm$^{-2}$~s$^{-1}$~arcsec$^{-2}$. 
The [S~{\sc ii}] $\lambda$6716 \AA\ to [S~{\sc ii}] $\lambda$6731 \AA\ line ratio free from nebular emission has been used to estimate the electronic density ($N_e$) of the material in the jet assuming a $T_\mathrm{e}$ of 10,000 K (Fig.~\ref{fig:megara}-a). 
The density of the jet is lower than that of the nebula \citep{Pottasch2008}, with values of $N_e$ in the range from the low-density regime $\approx$100~cm$^{-3}$ up to a few thousands cm$^{-3}$, for a median $N_\mathrm{e}$ of 380$\pm$160 cm$^{-3}$. 
The integrated values of $N_e$ for the different components of the jet reveal a density gradient, from $\simeq400$ cm$^{-3}$ for the inner jet, $\simeq150$ cm$^{-3}$ for the mid jet, and at the low-density regime $\lesssim100$ cm$^{-3}$ for the outer jet.

\begin{figure*}
\begin{center}
\includegraphics[width=0.475\linewidth]{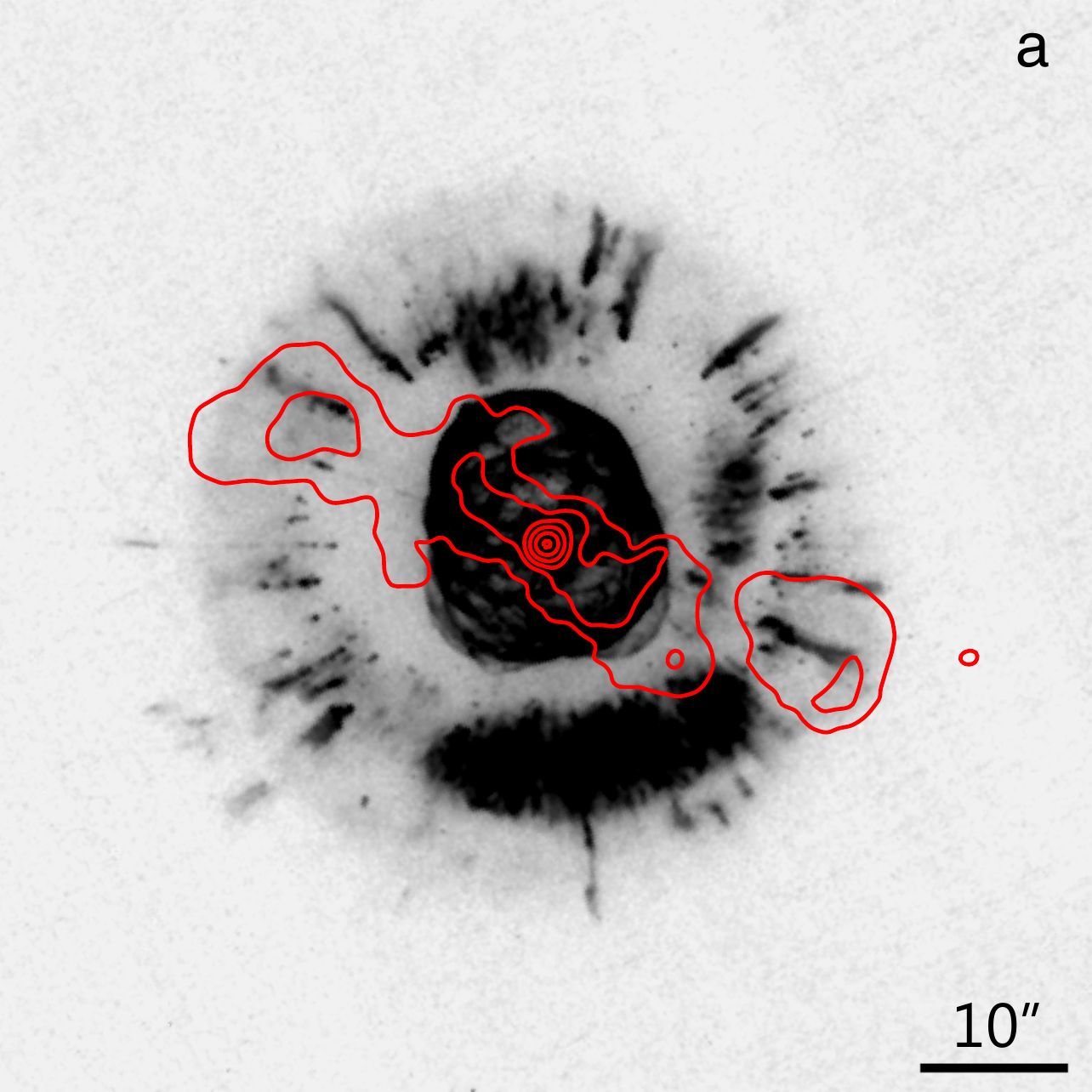} 
\hspace*{0.012cm}
\includegraphics[width=0.475\linewidth]{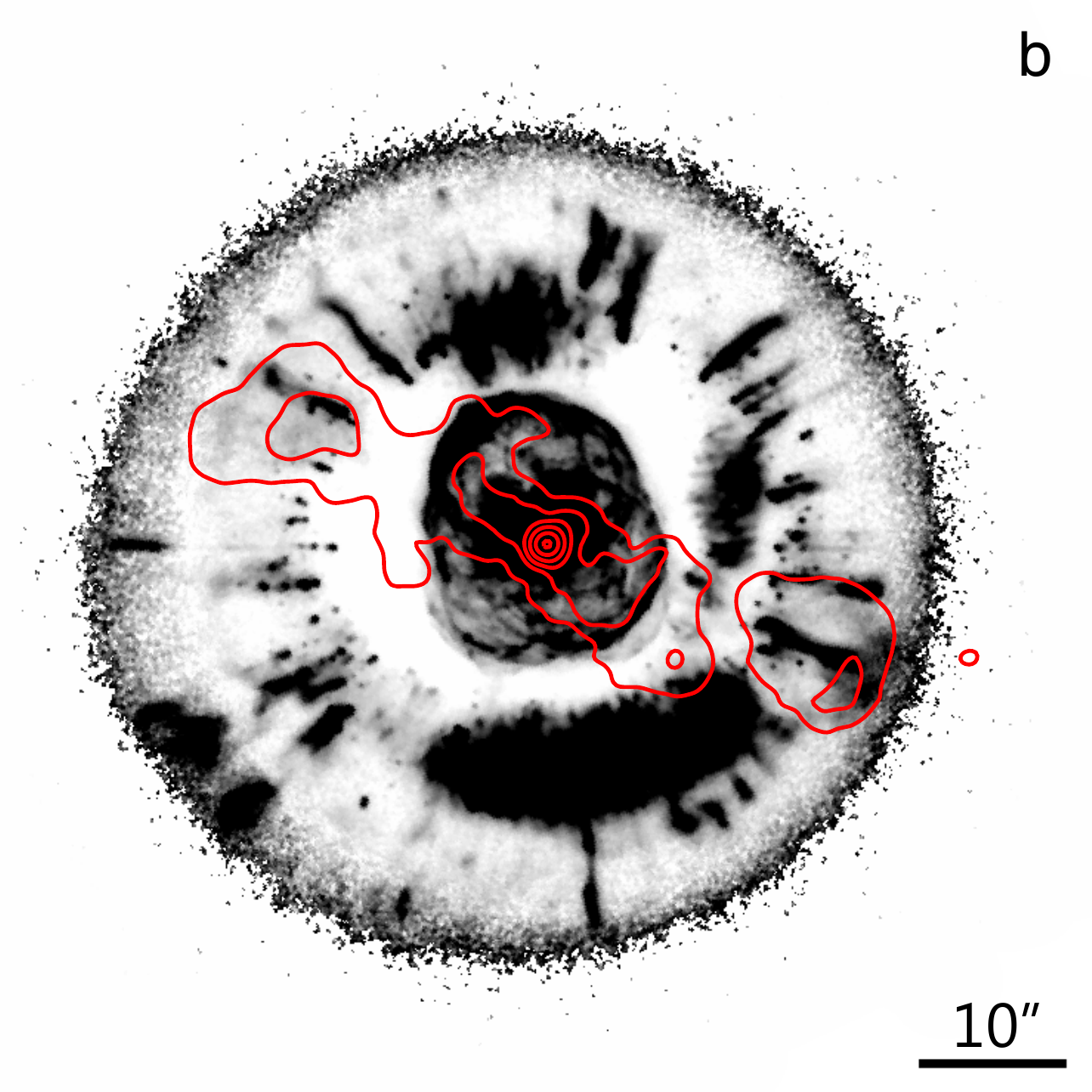} 
\caption{
\emph{HST} WFPC2 [S~{\sc ii}] image (left) and [N~{\sc ii}]/[O~{\sc iii}] ratio map (right). 
The contrast is selected to emphasize the faint emission from the mid jet. 
The red contours, extracted from the MEGARA GTC map of the jet in the [N~{\sc ii}] $\lambda$6584 \AA\ emission line (Fig.~\ref{fig:megara}-a), are also selected to highlight the mid jet emission projected onto the outer shell.
}
\vspace*{-0.35cm}
\label{fig:hst}
\end{center}
\end{figure*}

In retrospect, we have examined the \emph{HST} WFPC2 images of NGC\,2392 to search for the jet emission. 
The inner jet, projected onto the bright emission of the inner shell, cannot be detected, neither can the weak emission from the outer jet, but it is possible to identify in the [N~{\sc ii}] and [S~{\sc ii}] images and [N~{\sc ii}]/[O~{\sc iii}] and [S~{\sc ii}]/[O~{\sc iii}] ratio maps the emission from the mid jet projected onto the outer shell (Fig.~\ref{fig:hst}). 
The morphology of the mid jet shown by the GTC MEGARA data is confirmed in these
\emph{HST} images and ratio maps; the Eastern mid jet is revealed as a curving-blob, whereas the Western mid jet has a loop-like morphology with brighter emission at its tip.
We remark that the identification of the emission from the mid jet in \emph{HST} images and ratio maps has become possible only after a close comparison with the GTC MEGARA data, as the emission from the mid jet is basically indistinguishable from the numerous low-velocity patches of diffuse emission of the outer shell.

\subsection{Jet mass}

The ionized mass of a nebula can be derived using the relationship: 
\begin{equation}
M_\mathrm{i} = 11.06\times F({\rm H}\beta) \times d^2 \times T_\mathrm{e}^{0.88} \times N_\mathrm{e}^{-1} \;M_\odot
\end{equation}
if $N_\mathrm{e}$ is derived from density diagnostic emission lines and $F({\rm H}\beta)$ in units of $10^{-11}$ erg~cm$^{-2}$~s$^{-1}$, $d$ in kpc, and $T_\mathrm{e}$ in 10,000 K \citep{Pottasch1984}. 
It is possible to derive the intrinsic H$\beta$ flux from the observed H$\alpha$
flux adopting the recombination case B for an H$\alpha$ to H$\beta$ theoretical ratio of 2.86 \citep{Osterbrock2006} and correcting the reddening using a value of 0.23 for $c$(H$\beta$)\citep{Pottasch2008}. 
Since it is not possible to build a complete map of the surface brightness in the H$\alpha$ emission line, the [N~{\sc ii}] $\lambda$6584 \AA\ to H$\alpha$ emission line ratio has been measured in spectra extracted in a number of representative apertures and its average and dispersion values estimated to be $1.5\pm0.2$.

Accordingly, the intrinsic H$\beta$ flux is derived to be $(2.6\pm0.4)\times10^{-14}$ erg~cm$^{-2}$~s$^{-1}$.
Adopting the distance of $1.84\pm0.16$ kpc derived from the {\it Gaia} EDR3 parallax \citep{Gaia2020}
and assuming a $T_\mathrm{e}$ of 10,000~K, an ionized mass $M_\mathrm{i}$ of $(2.5\pm1.1)\times10^{-4} M_\odot$ is derived for the jet of NGC\,2392. 
The error bar includes the 1-$\sigma$ uncertainties in flux, distance, and electronic density.

\subsection{Jet inclination, velocity, age and mass loss rate}

Besides the jet mass, its inclination with the line of sight is a key parameter that would allow the determination of its space velocity, age, and mass-loss rate. 
In general, it would not be possible to derive the inclination of a jet unless additional assumptions were adopted.
As for the jet of NGC\,2392, its inclination angle has been proposed to be 7$^\circ$ assuming it has the same age of the inner shell \citep{Gieseking1985} and in the range $3.3^\circ$--$10^\circ$ assuming it shares the same tilt with the line of sight of the inner shell \citep{Odell1985}. 
The inner jet is found at an average PA on the plane of the sky $\sim$55$^\circ$ and then it twists to $\sim$70$^\circ$ in the mid jet (Fig.~\ref{fig:megara}-a). 
The jet is thus clearly misaligned with the inner shell whose symmetry axis PA has been estimated to be $20^\circ$--$25^\circ$ \citep{GarciaDiaz2012,Odell1990}, questioning the above assumptions on the similarity between the jet and inner shell inclinations.

\begin{figure}
\begin{center}
\includegraphics[bb=165 175 410 405,width=0.9\linewidth]{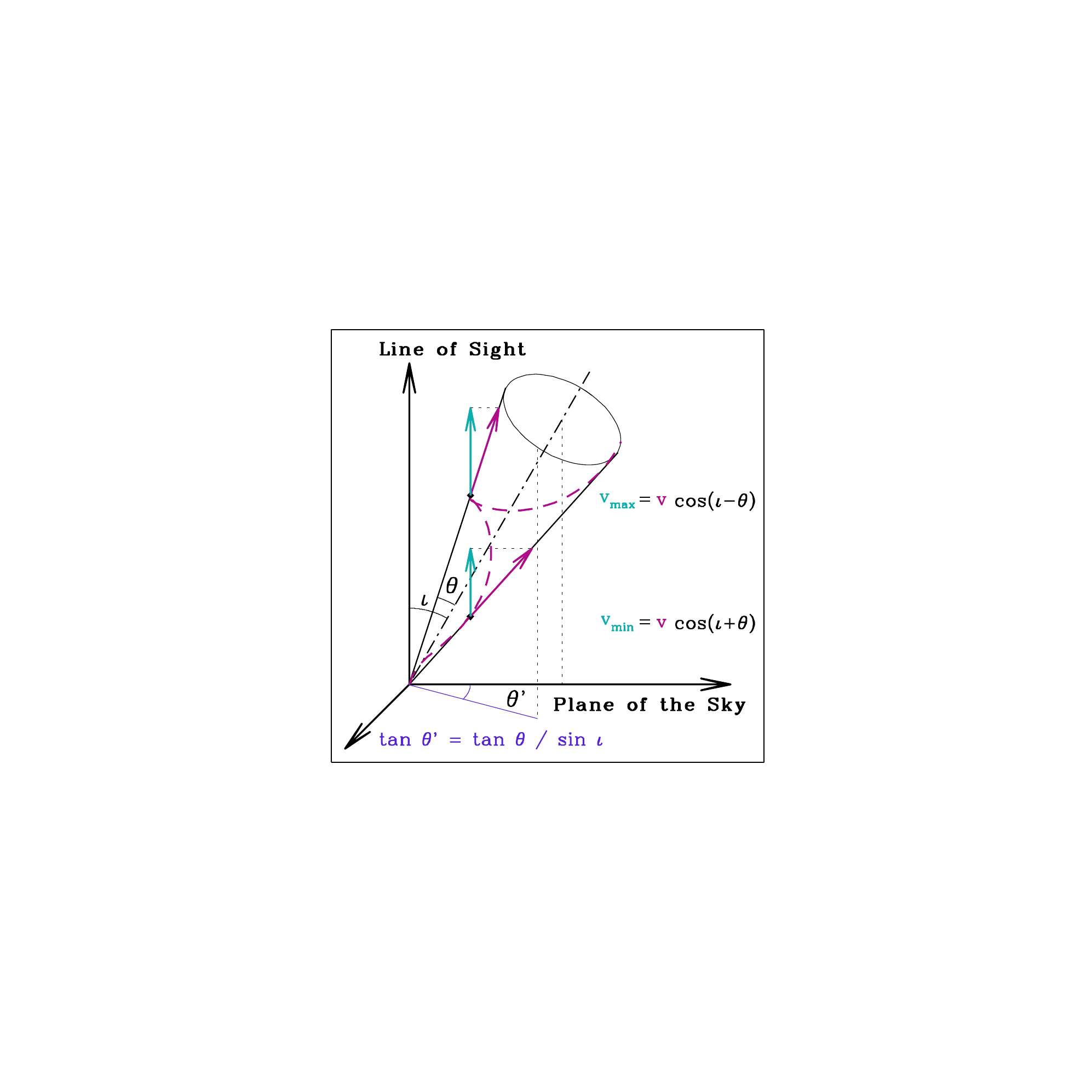} 
\caption{
Sketch of the geometry of a precessing jet with inclination of the symmetry axis of the precession cone with the line of sight $\iota$ and half-aperture angle of the precession cone $\theta$. 
The relationships between observed maximum ($v_{\rm max}$) and minimum ($v_{\rm min}$) radial velocities and projected half-aperture angle of the precession cone $\theta'$ with $\iota$, $\theta$ and the jet space velocity ($v$) are shown.
}
\vspace*{-0.35cm}
\label{fig:sketch}
\end{center}
\end{figure}

\begin{figure*}
\begin{center}
\includegraphics[bb=2 255 574 420,width=1.00\linewidth]{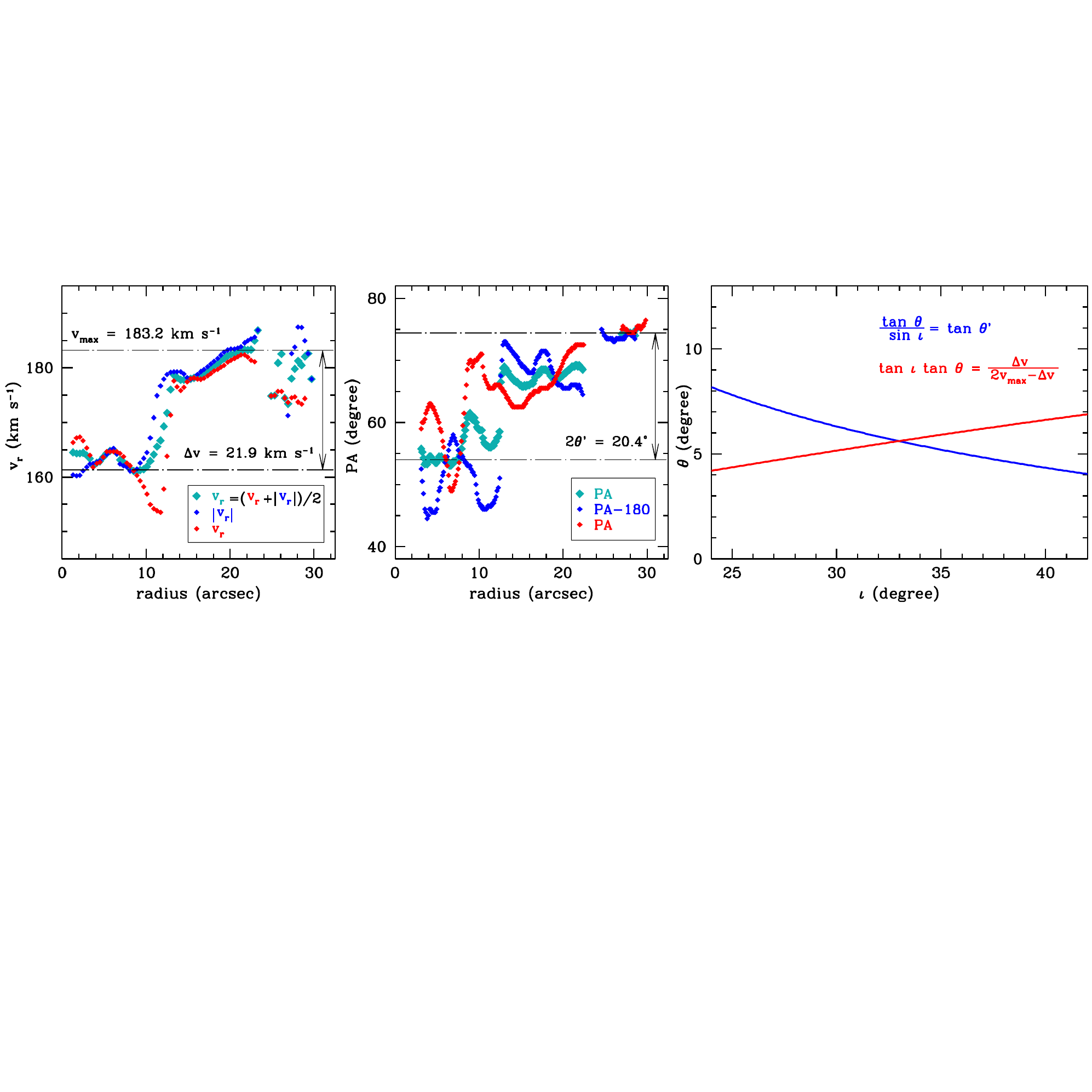} 
\caption{
(left) Variation of the radial velocity with respect to the systemic velocity with radius for the receding (red) and approaching (blue) components. 
The average of the absolute values of the receding and approaching velocities is also shown (turquoise). 
The maximum velocity $v_{\rm max}$ and its difference with the minimum velocity $\Delta v$ are shown. 
(center) Variation of the position angle of the receding (red) and approaching (blue) components. 
The average value of these two position angles is also shown (turquoise). 
The value of the projected half-aperture angle of the precession cone $\theta'$ is shown. 
(right) Variations of the inclination with the line of sight $\iota$ with the half-aperture angle of the precession cone $\theta$ of the jet of NGC\,2392 (solid lines) according to the relations (5) and (6) as labelled on the top-right of the
panel and the values of $v_{\rm max}$, $\Delta v$ and $\theta'$ obtained in panels a and b.
}
\vspace*{-0.35cm}
\label{fig:i-theta}
\end{center}
\end{figure*}

The S-shaped morphology of the jet and its correlated radial velocity variations (Fig.~\ref{fig:megara}-a and b) are consistent with the spatio-kinematic behavior of an episodic precessing jet. 
Following \citet{Guerrero1998}, a precessing jet can be described by the half-aperture angle of the precession cone ($\theta$), inclination angle of its symmetry axis with the line of sight ($\iota$) and space velocity ($v$). 
Since the approaching and receding components of the jet are found at different sides of the central star, it is inferred that the half-aperture angle of the precession cone $\theta$ is smaller than the inclination angle $\iota$ as assumed in Figure~\ref{fig:sketch}.
The maximum and minimum systemic radial velocities of the jet ($v_{\rm max}$ and $v_{\rm min}$) depend on the expansion velocity of the jet $v$ and on $\theta$ and $\iota$ as
\begin{equation}
    v_{\rm max} = v \cos(\iota - \theta)
\end{equation}
\begin{equation}
    v_{\rm min} = v \cos(\iota + \theta) 
\end{equation}
which can be expressed as
\begin{equation}
    v_{\rm max} - v_{\rm min} = \Delta v = 2 \, v \sin \iota \sin \theta.
\end{equation}
Since $v$ is not known but it can be substituted using equation (2), after some calculations we derive the following expression
\begin{equation}
\tan \iota \tan \theta = \frac{\Delta v}{2 \, v_{\rm max} - \Delta v}.
\end{equation}
The radial dependence of the systemic radial velocity of the jet ($v_r$) has been obtained by computing its average value at each radius from Figure~\ref{fig:megara}-b. 
The values of $v_r$ for the approaching and receding jet and its average value are shown in Figure~\ref{fig:i-theta}-left.
From this figure, values of 183.2 km~s$^{-1}$ and 21.9 km~s$^{-1}$ are obtained for $v_{\rm max}$ and $\Delta v$, respectively. 
The small value of the ratio $\Delta v/v_{\rm max}$ hints at small values of $\iota$ and $\theta$.

The projection of the half-aperture angle of the precession cone $\theta$ on the plane of the sky ($\theta'$) depends on $\theta$ and $\iota$ according to 
\begin{equation}
    \tan \theta' = \frac{\tan \theta}{\sin \iota}
\end{equation}
as can be inferred from Figure~\ref{fig:sketch}.  
The averaged PA at each radius of the jet has been computed from the surface brightness map in Figure~\ref{fig:megara}-a. 
The PA of the receding and approaching components of the jet and their average value are shown in Figure~\ref{fig:i-theta}-center, from where a value of 10.2$^\circ$ is derived for $\theta'$.

The values of $v_{\rm max}$ and $\Delta v$ define the black curve in Figure~\ref{fig:i-theta}-right. 
Similarly, the value of $\theta'$ defines the purple curve in Figure~\ref{fig:i-theta}-right. 
An inclination angle of the jet $\iota$ of 33.0$^\circ$ and half-aperture angle of the precession cone $\theta$ of 5.6$^\circ$ are defined where these two curves cross. 
According to equation (2), the space velocity of the jet is derived to be 206 km~s$^{-1}$.

\begin{figure*}
\begin{center}
\includegraphics[bb=30 120 605 330,width=0.90\linewidth]{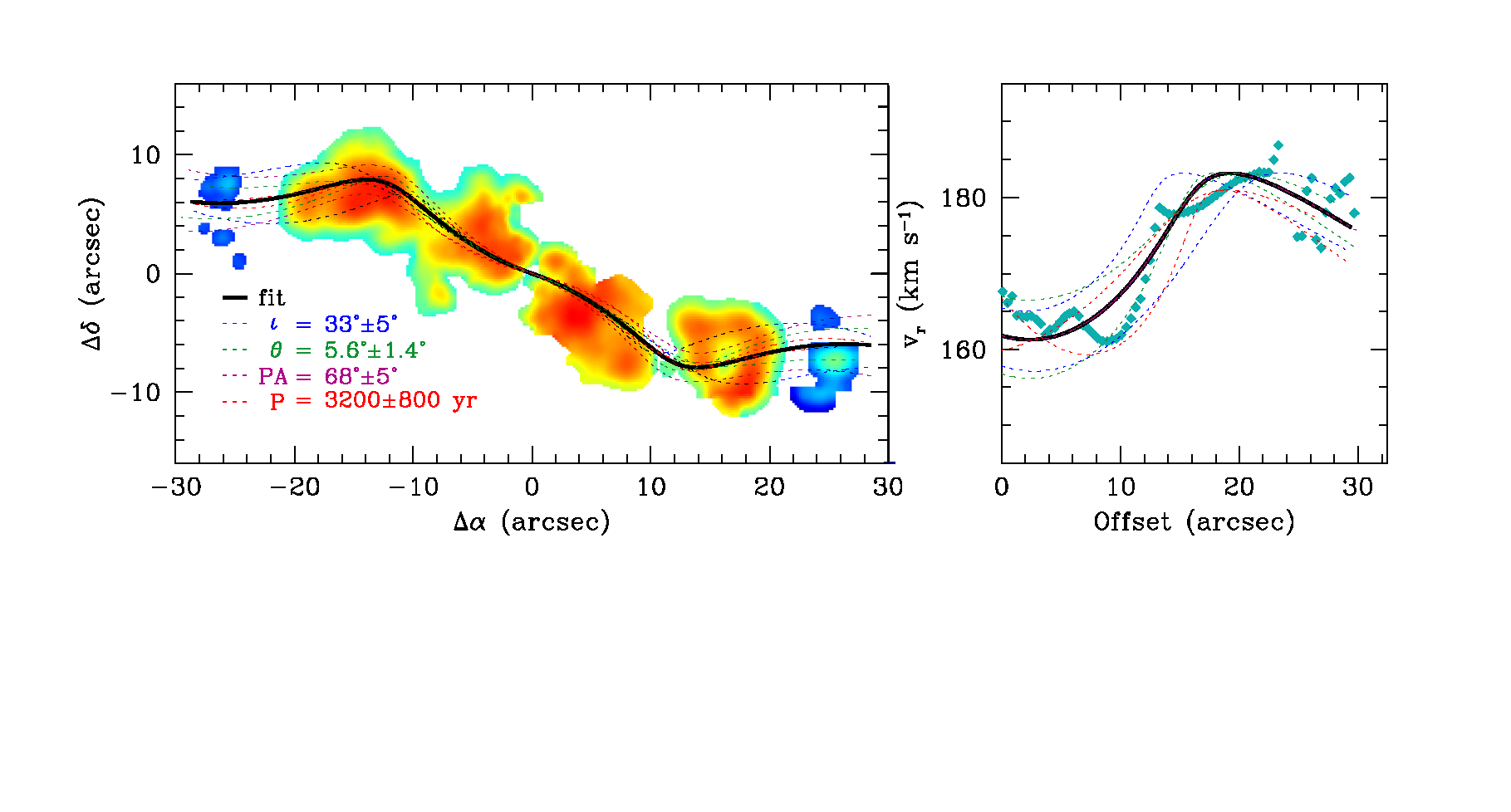} 
\caption{
Best-fit of the spatio-kinematic properties of the precessing jet sketched in Figure~\ref{fig:sketch} to the observed [N~{\sc ii}] $\lambda$6584 \AA\ surface brightness distribution presented in Figure~\ref{fig:megara}-a (left) and to the azimuthally averaged radial velocity of the approaching and receding jet components presented in Figure~\ref{fig:i-theta}-center (right). 
The best-fit (thick solid line) assumes $\iota$ to be 33.0$^\circ$, $\theta$ to be 5.6$^\circ$, and $v$ to be 206 km~s$^{-1}$ as derived in the text.  
The best fit implies a precession period (P) of 3200 yr, an age ($\tau$) of 2400 yr, and a PA on the sky of 68$^\circ$. 
The dotted lines correspond to models allowing variations of $\pm5^\circ$ in $\iota$, $\pm$1.4$^\circ$ in $\theta$, $\pm$5$^\circ$ in PA, and $\pm$800 yr in P, where $v$ and $\tau$ were adjusted accordingly.
}
\vspace*{-0.35cm}
\label{fig:fit}
\end{center}
\end{figure*}

The S-shaped morphology of the jet and its velocity variations have been subsequently modeled assuming a continuous ejection of a precessing jet with the velocity $v$, inclination $\iota$ and aperture angle $\theta$ given above. 
This simple analytical model indeed provides a reasonable fit to the observed morphology (Fig.~\ref{fig:fit}-left) and kinematics (Fig.~\ref{fig:fit}-right) of the jet of NGC\,2392 --taking into account their notable complexity and the episodic nature of the ejection-- for a precession period of 3200 yr and an age of 2400 yr at the adopted \emph{Gaia} distance of 1.84 kpc.
The full linear span of the jet would be 1.0 pc, i.e., $\approx$ 2.3 times larger than the nebular diameter of 0.43 pc derived from its angular size. 
Its mass-loss rate would be $1.0\times10^{-7} M_\odot$~yr$^{-1}$ and the mechanical luminosity 0.4 $L_\odot$, a tiny fraction of the 3000--6000 $L_\odot$ stellar luminosity \citep{Herald2011}.

Figure~\ref{fig:fit} is additionally overlaid with variations of the different parameters still producing reasonable fits to the observations. 
These variations define the uncertainty in jet parameters, but they can be rather envisaged as intrinsic velocity or direction fluctuations in the launch of the jet or variations caused by its interaction with the nebular material that are not considered in the simplistic model used here to describe its morphology and kinematics.

\section{Discussion} 
\label{sec:discussion}

\subsection{The jet and the nebula}

The brightening of the inner jet at the inner shell rim (Fig.~\ref{fig:img}-left) suggests the interaction of the jet with the inner shell. 
Projection effects can be important, though, as the inner jet has a full linear span $\simeq$0.4 pc, whereas the maximum size of the inner shell, as projected in the plane of the sky, is only $\simeq$0.2 pc.  
The size of the inner shell along the line of sight is most likely considerably larger, as suggested by the model presented by \citet{GarciaDiaz2012}, may be comparable to the inner jet extent.  
The similarity between the expansion velocity of the inner shell, $\simeq$120 km~s$^{-1}$ \citep{GarciaDiaz2012}, and that of the inner jet at the location of its rim, $\simeq$150 km~s$^{-1}$, suggests that the inner shell is being dragged by the jet.

It is also worthwhile noting the correspondence between the mid jet morphology and some cometary knots of the outer shell.  
The loop-like feature of the Western mid jet seems to surround a cometary knot and the Eastern mid jet curves as it seems to reach a cometary knot.  
These morphological correspondences might be suggestive of the interaction of the jet with nebular material, but we note that those cometary knots have been suggested to be close to the plane of the sky, whereas the jet is only 33$^\circ$ apart from the line of sight.

A comprehensive study of the detailed spatio-kinematic properties of the inner and outer shells of NGC\,2392 and its cometary knots to investigate the possible jet-nebula interactions will be the goal of a subsequent work (Rechy-Garc\'\i a et al., in preparation).

\subsection{A unique on-going jet among PNe}

The close-up view of the jet emission in Figure~\ref{fig:img}-left shows that, within the spatial resolution provided by the present observations, the jet of NGC\,2392 emanates from its central star, confirming previous suggestions based on the emergence of the jet emission from the stellar continuum in high-dispersion echelle spectroscopic observations \citep{Gieseking1985,GarciaDiaz2012}.  
On these grounds, the high-ionization polar stellar wind proposed for its central star \citep{Prinja2014} could be interpreted as the onset of the jet rather than an isotropic stellar wind. 
Indeed, the terminal velocity of such stellar wind $\approx$ 300 km~s$^{-1}$ is abnormally low for central stars of PNe, but close to the velocity of the jet.

The launch mechanism of the jet of NGC\,2392 is thus presently active, i.e., it is an on-going jet \citep{Miszalski2019}. 
This is also the case of the late-AGB jets in BD+46$^\circ$442 and IRAS\,19135$+$3937 \citep{Bollen2019}. 
The mass-loss rates of those jets is $(0.5-1)\times10^{-6}$~M$_\odot$ yr$^{-1}$ and $(1.3-4)\times10^{-6}$~M$_\odot$~yr$^{-1}$, respectively \citep{Bollen2020}.
Adopting a ratio of ejected mass-loss rate to accretion rate for the jet in the range 0.01--0.3 for NGC\,2392, similar to that adopted by \citet{Bollen2020}, the accretion rate would be in the range $4\times10^{-7}$–10$^{-5}$~M$_\odot$~yr$^{-1}$. 
The high-regime of accretion of late-AGB jets sustained by the heavy stellar mass-loss at this phase is apparently not operating in NGC\,2392, which is a mature PN.

This seems natural, as once the central star of a PN enters the post-AGB phase and its heavy mass-loss ceases, the accretion onto a companion and the jet formation can be expected to come to an end too. 
Indeed, jets are found to be mostly coeval with their PNe \citep{Guerrero2020}. 
Only the jets of the post-common envelope (post-CE) PNe Hb\,4, NGC\,6337 and NGC\,6778 \citep{Derlopa2019,GarciaDiaz2009,Tocknell2014} have been proposed to originate during the post-AGB phase. 
These post-CE jets would be fed by the re-accretion of nebular gas or material from the remains of a circumbinary disk onto the primary \citep{Soker1994}.

Besides the case of Hen\,2-90 \citep{SN2000,Guerrero2001}, which might be a symbiotic star rather than a PN \citep{SK2001}, all the jets in PNe, including those in post-CE systems, are spatially detached from their central stars, with notable spatial gaps from them to the innermost regions of the jets.  
It is then implied that the jet collimation in PNe ceased hundreds to thousands of years ago, making the jet in NGC\,2392 the only on-going one observed in a mature PN.

\subsection{NGC~2392 in the ecosystem of PNe}

NGC\,2392 has been proposed to be an analogue to NGC\,6543 and NGC\,7009 with a pole-on orientation \citep{GarciaDiaz2012}. 
More naturally, the large jet-to-nebula size ratio of NGC\,2392 derived above, $\approx$ 2.3, and the fast expansion velocity and S-shaped morphology of its jet would make it rather an analogue to Fleming\,1, the archetype of PNe with bipolar, rotating, episodic jets \citep[BRET,][]{Lopez1993,Palmer1996}. 
The physical structure of NGC\,2392, with its ring of high-density knots \citep{Odell1990} and long precessing jet is also reminiscent of IPHASX\,J194359.5$+$170901 \citep{Corradi2011}. 
Interestingly, both Fleming\,1 and IPHASX\,J194359.5$+$170901 harbour post-CE binary systems \citep{Boffin2012,Miszalski2013}. 
Their jets have been suggested to pre-date their formation \citep{Tocknell2014}, which may be the case of NGC\,2392 as well given the age of $1140/d$ yr~kpc$^{-1}$  proposed for the inner shell \citep{Odell1985}. 
NGC\,2392 would then be a young twin of Fleming\,1 and IPHASX\,J194359.5$+$170901 caught in the short time lapse between the PN formation and the jet extinction.

Finally, it is worth noting that the central star of NGC\,2392 has recently been recognized to be in a binary system \citep{Miszalski2019}. 
Its orbital period is yet unsettled, as periodic radial velocity variations of 3 hours \citep{Prinja2014} and 1.9 days \citep{Miszalski2019} and a hard X-ray emission modulation of 6 hours \citep{Guerrero2019} have been reported.
The companion of the central star of NGC\,2392 has been proposed to be a hot white dwarf as its effective temperature $\simeq$43,000 K \citep[][]{Mendez2012} cannot explain its high nebular excitation, being notably discrepant from the higher He~{\sc ii} Zanstra temperature \citep{Heap1977,Pottasch2008,Miszalski2019}.  
This would make the central star of NGC\,2392 one of the few double-degenerate ones as Fleming\,1 itself or Hen\,2-428 \citep{Boffin2012,SantanderGarcia2015,Reindl2020}. 
The present-day launch and collimation of a jet in NGC\,2392 strongly supports the presence of a binary system and an accretion disk around a compact source at its heart. 
The low mass-loss rate of its central star, $(3-5)\times10^{-8} M_\odot$~yr$^{-1}$, is not capable however to feed the material in the jet \citep{Herald2011}, which would be rather accreted from a circumbinary disk remnant of a CE episode late in the AGB phase. 
Indeed, the peculiar chemical abundances of the central star of NGC\,2392 have been suggested to originate in such a CE phase \citep{Mendez2012}. 
The central star of NGC\,2392 would be in its final transition towards a double-degenerate binary system, when the second component of the system evolves into the white dwarf stage ejecting its envelope to form a PN-like shell. 
It is of upmost importance to determine the real orbital period and masses of the double-degenerate binary system in NGC\,2392 
to assess its possible fate as a Type Ia supernova.

\section{Summary} 
\label{sec:summary}

The jet in NGC\,2392 was the first fast collimated outflow ever detected in a PN, yet a direct image of this jet was lacking.  
This peculiar situation was caused not only by the low surface brightness of the jet projected onto the much brighter nebular emission, but also by the high expansion velocity of the inner shell, close to the velocity of the jet in the innermost regions of NGC\,2392. 
The lack of an exact description of the morphology and kinematics of the jet in NGC\,2392 impeded a clear understanding of its role in the formation and shaping of this PN.

To overcome these difficulties, we have obtained high-dispersion IFS observations of NGC\,2392 using MEGARA and its high-resolution VPH665-HR grism at the 10.4~m GTC in the spectral range of the [N\,{\sc ii}]~$\lambda\lambda$6548,6584~\AA, H$\alpha$, and [S\,{\sc ii}]~$\lambda\lambda$6716,6731~\AA\ emission lines. 
The 2D spectra presented here, with their spectral resolution $\approx$ 16 km~s$^{-1}$, unprecedentedly resolves the [N\,{\sc ii}] and [S\,{\sc ii}] emission of the jet from that of the nebula to build maps of their surface brightness, velocity, velocity width, and electronic density and to derive its total mass.

The jet of NGC\,2392 is found to consist mostly of two large blobs and a few fainter knots emerging from the central star and extending outside the edge of the outer shell.  
The S-shaped morphology of the jet and its correlated velocity variations are highly suggestive of a precessing jet. 
Adopting a model of the continuous ejection of a precessing jet, an inclination angle with the line of sight of 33.0$^\circ$ and a space velocity of 206 km~s$^{-1}$ are derived.  
Accordingly, the linear span of the jet of NGC\,2392 is found to be 1.1 pc, its age $\approx$ 2600 yr, and its mass-loss rate $\approx 1.1\times10^{-7} M_\odot$~yr$^{-1}$.

The linear size of the jet of NGC\,2392 is thus $\approx$ 2.3 times larger than the nebula.  
The jet pierces through the bright inner shell, brightening at its rim.  
The expansion velocity of the jet at these locations is close to that of the inner shell, suggesting that the jet powers the abnormally high speed expansion of the inner shell of NGC\,2392 rather than its feeble stellar wind.  
The jet of NGC\,2392 may be the primary mechanism for the shaping and expansion of its inner shell.

Very importantly, the jet in NGC\,2392 is found to be currently active, i.e., it is being launched right now.  
This is unique among the many collimated outflows found in PNe, which have been hitherto described as fossil evidence of previous ejections.  
The hard X-ray emission from the central star of NGC\,2392 and the high excitation of the nebula strongly support a white dwarf as the binary companion of the central star. 
The jet in NGC\,2392 would arise from an accretion disk around this compact companion fed by the circumbinary disk of a previous common-envelope phase.

\acknowledgments

MAG amd BMM acknowledge support of the Spanish Ministerio de Ciencia, Innovaci\'on y Universidades (MCIU) grant PGC2018-102184-B-I00.
MAG, SC and BMM acknowledge financial support from the State Agency for Research of the Spanish MCIU through the “Center of Excellence Severo Ochoa” award
to the Instituto de Astrof\'\i sica de Andaluc\'\i a (SEV-2017-0709).
GR-L acknowledges support from CONACyT (grant 263373) and PRODEP (Mexico).
JSRG and VMAGG acknowledge support from the Programa de Becas posdoctorales funded by Direcci\'on General de Asuntos del Personal Acad\'emico (DGAPA) of the Universidad Nacional Aut\'onoma de M\'exico (UNAM). 
JAT acknowledges funding by DGAPA UNAM PAPIIT project IA100720 and the Marcos Moshinsky Foundation (Mexico). 
XF acknowledges support by the Strategic Priority Research Program of Chinese Academy of Sciences, Grant No.\ XDB 41000000.

The authors deeply appreciate Profs.\ Bruce Balick and You-Hua Chu for a critical reading of the manuscript.

This work has made use of data from the European Space Agency (ESA) mission {\it Gaia} (\url{https://www.cosmos.esa.int/gaia}), processed by the {\it Gaia} Data Processing and Analysis Consortium (DPAC, \url{https://www.cosmos.esa.int/web/gaia/dpac/consortium}). 
Funding for the DPAC has been provided by national institutions, in particular the institutions participating in the {\it Gaia} Multilateral Agreement.

\facilities{Gran Telescopio de Canarias (MEGARA), \emph{Hubble Space Telescope} (WFPC2).}

\software{megaradrp v0.10.1, megarasss2cube, IDL {\sc mpfitexpr}, JMAPLOT.}


\end{document}